\documentclass[fleqn,twoside]{article}
\usepackage[headings]{espcrc2}

\readRCS
$Id: espcrc2.tex,v 1.2 2004/02/24 11:22:11 spepping Exp $
\usepackage{graphicx}
\usepackage[figuresright]{rotating}

\makeatletter
\def\nl{\nonumber\\}
\makeatother

\def\beq{\begin{equation}}
\def\eeq{\end{equation}}
\def\beqar{\begin{eqnarray}}
\def\eeqar{\end{eqnarray}}
\def\barr#1{\begin{array}{#1}}
\def\earr{\end{array}}
\def\nn{\nonumber}

\def\ga{\gamma}
\def\de{\delta}

\def\Si{\Sigma}

\def\refeq#1{\mbox{(\ref{#1})}}
\def\citere#1{\mbox{Ref.~\cite{#1}}}
\def\citeres#1{\mbox{Refs.~\cite{#1}}}

\newcommand{\ri}{{\mathrm{i}}}
\newcommand{\rT}{{\mathrm{T}}}
\newcommand{\rw}{\mathswitchr w}

\newcommand{\ord}{\mathswitch{{\cal{O}}}}
\newcommand{\Oa}{\mathswitch{{\cal{O}}(\alpha)}}
\newcommand{\Oaa}{\mathswitch{{\cal{O}}(\alpha^2)}}
\newcommand{\Oaaa}{\mathswitch{{\cal{O}}(\alpha^3)}}

\def\mathswitchr#1{\relax\ifmmode{\mathrm{#1}}\else$\mathrm{#1}$\fi}
\newcommand{\Pf}{\mathswitch  f}
\newcommand{\PW}{\mathswitchr W}
\newcommand{\PZ}{\mathswitchr Z}
\newcommand{\PH}{\mathswitchr H}
\newcommand{\Pep}{\mathswitchr {e^+}}
\newcommand{\Pem}{\mathswitchr {e^-}}

\def\mathswitch#1{\relax\ifmmode#1\else$#1$\fi}
\newcommand{\MW}{\mathswitch {M_\PW}}
\newcommand{\MZ}{\mathswitch {M_\PZ}}
\newcommand{\GW}{\Gamma_{\PW}}
\newcommand{\GZ}{\Gamma_{\PZ}}

\newcommand{\sw}{\mathswitch {s_\rw}}
\newcommand{\cw}{\mathswitch {c_\rw}}

\def\ie{i.e.\ }
\def\eg{e.g.\ }

\def\Re{\mathop{\mathrm{Re}}\nolimits}
\def\Im{\mathop{\mathrm{Im}}\nolimits}

\def\draftdate{\relax}
\def\mda{\relax}
\def\mua{\relax}
\def\mla{\relax}
\def\draft{
\def\thtystars{******************************}
\def\sixtystars{\thtystars\thtystars}
\typeout{}
\typeout{\sixtystars**}
\typeout{* Draft mode!
         For final version remove \protect\draft\space in source file *}
\typeout{\sixtystars**}
\typeout{}
\def\draftdate{\today}
\def\mua{\marginpar[\boldmath\hfil$\uparrow$]%
                   {\boldmath$\uparrow$\hfil}%
                    \typeout{marginpar: $\uparrow$}\ignorespaces}
\def\mda{\marginpar[\boldmath\hfil$\downarrow$]%
                   {\boldmath$\downarrow$\hfil}%
                    \typeout{marginpar: $\downarrow$}\ignorespaces}
\def\mla{\marginpar[\boldmath\hfil$\rightarrow$]%
                   {\boldmath$\leftarrow $\hfil}%
                    \typeout{marginpar: $\leftrightarrow$}\ignorespaces}
\def\Mua{\marginpar[\boldmath\hfil$\Uparrow$]%
                   {\boldmath$\Uparrow$\hfil}%
                    \typeout{marginpar: $\uparrow$}\ignorespaces}
\def\Mda{\marginpar[\boldmath\hfil$\Downarrow$]%
                   {\boldmath$\Downarrow$\hfil}%
                    \typeout{marginpar: $\downarrow$}\ignorespaces}
\def\Mla{\marginpar[\boldmath\hfil$\Rightarrow$]%
                   {\boldmath$\Leftarrow $\hfil}%
                    \typeout{marginpar: $\leftrightarrow$}\ignorespaces}
\overfullrule 5pt
\oddsidemargin -15mm
\marginparwidth 29mm
}

\newcommand{\cmws}{\mu^2_\PW}
\newcommand{\cmzs}{\mu^2_\PZ}

\newcommand{\csw}{\mathswitch {s_\rw}}
\newcommand{\ccw}{\mathswitch {c_\rw}}
\newcommand{\cZ}{\mathcal{Z}}

\newcommand{\AmS}{{\protect\the\textfont2
  A\kern-.1667em\lower.5ex\hbox{M}\kern-.125emS}}

\title{The complex-mass scheme for perturbative calculations with
  unstable particles}

\author{A. Denner\address{Paul Scherrer Institut, 
        W\"urenlingen und Villigen,
        CH-5232 Villigen PSI, Switzerland}%
        \thanks{supported in part by the Swiss National Science
          Foundation},
        S. Dittmaier\address{Max-Planck-Institut f\"ur Physik 
(Werner-Heisenberg-Institut), D-80805 M\"unchen, Germany},
}
       
\runtitle{The complex-mass scheme  for perturbative calculations with
  unstable particles}
\runauthor{A. Denner, S. Dittmaier}

\begin{document}

\begin{abstract}
  Perturbative calculations with unstable particles require the
  inclusion of their finite decay widths. A convenient, universal
  scheme for this purpose is the complex-mass scheme. It fully
  respects gauge-invariance, is straight-forward to apply, and has been
  successfully used for the calculation of various tree-level
  processes and of the electroweak radiative corrections to
  $\Pep\Pem\to4\Pf$ and $\PH\to4\Pf$.  \vspace{1pc}
\end{abstract}

\maketitle

\section{Introduction}

The upcoming Large Hadron Collider (LHC) and the planned International
Linear Collider (ILC) will allow for important tests of the Standard
Model and searches for physics beyond. These investigations are based
on scattering and decay processes with many external particles.  A
decent exploitation of these experiments requires adequate
theoretical tools. At the LHC, at least the NLO QCD corrections must
be taken into account even for background processes in order to obtain
useful predictions, and for many signal processes also the NNLO QCD
and NLO electroweak corrections are needed. At the ILC perturbative
corrections are mandatory for all precision investigations. 

The calculation of NLO corrections for processes with more than two
particles in the final state poses several problems. First, the large
number and complexity of Feynman diagrams
require methods to deal with and to simplify large algebraic expressions.
Second, the numerically stable evaluation of the loop integrals
necessitates suitable techniques, and the integration over the
complicated many-particle phase space, involving many singularities,
needs appropriate tools.  On top of this the appearance of internal
unstable particles provides additional complications.

The description of resonances in perturbation theory requires a Dyson
summation of self-energy insertions. This leads to a mixing of
perturbative orders and, if done carelessly, easily compromises
gauge invariance \cite{Berends:1969nt,Argyres:1995ym}.  Therefore, the
proper introduction of finite-width effects is non trivial.  For
lowest-order predictions several solutions have been described
\cite{Argyres:1995ym,Denner:1999gp,Aeppli:1993cb,Beenakker:1996kn,Stuart:1991xk,Aeppli:1993rs,Baur:1995aa,Passarino:1999zh,Beenakker:1999hi,Beneke:2003xh}.
For the evaluation of radiative corrections in the presence of
resonances, a pole expansion
\cite{Aeppli:1993cb,Stuart:1991xk,Aeppli:1993rs} has
been used in the past. 
It provides a gauge-invariant answer, but
restricts the validity of the result to the resonance region only and
is not reliable in threshold regions. 
Threshold regions might be covered by effective field theories
\cite{Beneke:2003xh} where a pole expansion is combined with a
dedicated expansion around the threshold.
Obviously, a scheme
would be desirable that is universally valid and easy to implement.
This is provided by the complex-mass scheme (CMS).  It
constitutes a viable, unified description that is applicable in the
complete phase space and does not require any matching between
different treatments for different regions.

\section{The complex-mass scheme at tree level}

In the CMS, which  was introduced in \citere{Denner:1999gp}
for lowest-order calculations, the W- and Z-boson
masses are consistently considered as complex quantities, defined as
the locations of the poles in the complex $k^2$ plane of the
corresponding propagators with momentum $k$.  Gauge invariance is
preserved if the complex masses are introduced everywhere in the
Feynman rules, in particular also in the definition of the weak mixing
angle,
\beq\label{eq:compl-mixing} 
\cos^2\theta_\PW \equiv \ccw^2 = 1-\csw^2 =
{\cmws}/{\cmzs},
\eeq
which is derived from the ratio of the complex mass squares
of the gauge bosons,%
\beqar\label{eq:complex-masses} 
\cmws&=& \MW^2 - \ri\MW\GW, \quad \cmzs= \MZ^2 - \ri\MZ\GZ.  
\eeqar 
All relations that do not involve complex conjugation, such as Ward
or Slavnov--Taylor identities, remain valid, because the
gauge-boson masses are modified only by an analytic continuation.  As
a consequence the resulting matrix elements are gauge-parameter
independent, and unitarity cancellations are respected. These
properties hold order by order in perturbation theory, despite the
fact that some higher-order contributions are incorporated in the
complex masses.

While necessary in the resonant propagators, the consistent
introduction of complex masses introduces spurious terms in other
places, as \eg in the weak mixing angle \refeq{eq:compl-mixing}.  When
using the CMS at tree level, which amounts to replacing the real
gauge-boson masses by the complex masses \refeq{eq:complex-masses} and
the weak mixing angle by \refeq{eq:compl-mixing} in tree-level
amplitudes, the spurious terms are of order $\ord(\GW/\MW)=\Oa$
relative to the lowest-order 
(both in resonant and non-resonant regions).

\section{The complex-mass scheme at one loop}

The generalization of the CMS to the one-loop level
was proposed in \citere{Denner:2005fg}. The complex masses are
introduced directly at the level of the Lagrangian by splitting the
real bare masses into complex renormalized masses and complex
counterterms. 
Complex masses are not only introduced for the gauge bosons but for
all unstable particles such as Higgs bosons and top quarks.
This scheme has the
following properties:
{\leftmargini 0ex
\begin{itemize}
\itemindent 3ex
\labelwidth 2ex
\itemsep 0pt\parsep0pt
\item The Lagrangian yields Feynman rules with complex masses and
  counterterms with which perturbative calculations can be performed
  as usual.  Since the bare Lagrangian is not changed
  at all, but only its perturbative expansion is rearranged, no double
\looseness-1
  counting of terms occurs.
\item For each unstable particle, the real bare mass is split
  into a complex renormalized mass and a complex counterterm.  The
  imaginary part of the renormalized mass becomes part of the free
  propagator, while the imaginary part of the counterterm becomes part
  of a counterterm vertex.  As in ordinary renormalization, the former
  term is resummed but the latter is not.  Independently of the
  imaginary part that is added and subtracted,%
  \footnote{Such an idea was already proposed in
    \citere{Stuart:1990vk}.}  
  this procedure does not spoil the
  algebraic relations that govern gauge invariance, and unitarity
  cancellations are respected order by order.
\item Performing an $\Oa$ calculation in the CMS yields $\Oa$ accuracy
  everywhere in phase space provided the width that enters in the
  resonant propagators via the complex mass is calculated including at
  least $\Oa$ corrections. This is evident away from the resonances,
  where one could expand in terms of the width, thus recovering the
  usual perturbative expansion. In the resonance region, where the
  resonant contributions dominate, both the prefactors of the resonant
  propagators and the resonant propagators themselves are taken into
  account in $\Oa$, and our results differ by $\Oaa$ terms from a
  leading pole approximation where this is applicable. Thus, any
  spurious terms are of order $\Oaa$.
\item The CMS requires one-loop integrals with
  complex internal masses.  The IR-singular integrals can be found in
  \citere{Beenakker:1990jr}.  The non-IR-singular 2-point and 3-point
  functions can be easily obtained by analytical continuation of the
  results in \citere{'tHooft:1979xw}.  The 4-point integrals
  necessary for $\Pep\Pem\to4\Pf$ have been obtained by analytic
  continuation of the results of \citere{Denner:1991qq}.

\end{itemize}
}

Introducing complex masses and couplings seems to violate unitarity.
Obviously, the Cutkosky cutting equations \cite{Cutkosky:1960sp} are
no longer valid, and unitarity cannot simply be proven order by order
anymore.  However, since we do not modify the bare Lagrangian, the
unitarity-violating terms are of higher order, \ie of $\Oaa$ in an
$\Oa$ calculation.  Moreover, this unitarity violation cannot be
enhanced, because all Ward or Slavnov--Taylor identities are exactly 
preserved. In this
respect one should also mention that unstable particles should be
excluded as external states, and only the $S$-matrix connecting stable
particle states needs to be unitary, as has already been pointed out
by Veltman in the sixties \cite{Veltman:1963th}.  Of course, before
the CMS can be viewed as a rigorous procedure to define a renormalized
quantum field theory it has to be clarified whether one can directly
prove unitarity order by order in this formalism. In particular, it is
an interesting question whether one can construct modified cutting
\looseness -1
equations in the CMS.

\section{Complex renormalization}

The consistent introduction of complex masses in loop calculations
necessitates the formulation of an appropriate renormalization
prescription. To this end, we generalize the on-shell renormalization
scheme formulated in \citeres{Denner:1993kt,Aoki:1980ix,Denner:1994xt}
at the one-loop level in a straight-forward way.
A generalization to higher orders should be
possible.

For illustration here we treat only the renormalization of the W-boson
self-energy. The complete one-loop renormalization in the CMS was
presented in \citere{Denner:2005fg}.

The complex renormalized masses and mass counterterms result from a
splitting of the real bare masses squared,
\beq
M_{\PW,0}^2=\cmws+\de\cmws, 
\eeq
where 
bare quantities are indicated by a subscript 0.  Similarly, splitting
the bare W-boson field,
\beq
\textstyle
W_{0}^{\pm}   =  
\left(1+\frac{1}{2}\delta \cZ_{W}\right) W^{\pm} , 
\eeq
into a
complex field renormalization constant $\de \cZ_{W}$ and the
renormalized field
implies that the bare and renormalized fields have different phases.
We stress that 
$\de \cZ_{W}$ applies to
both the $W^+$ and $W^-$ field, \ie the imaginary part of $\de
\cZ_{W}$ is fixed by the renormalization condition and does not change
sign when going from the $W^+$ to the $W^-$ field.
As a consequence, the renormalized Lagrangian, \ie the Lagrangian in
terms of renormalized fields without counterterms, is not hermitian,
but the total Lagrangian (which is equal to the bare Lagrangian) of
course is.  

The renormalized
(indicated by the hat) transverse ($\rT$) W-boson self-energy reads
\beq\label{eq:ren-se}
\hat\Si^{W}_{\rT}(k^2) = \Si^{W}_{\rT}(k^2) - \de\cmws 
+(k^2-\cmws)\de \cZ_{W}.
\eeq
Compared to \citere{Denner:1993kt}
the renormalized on-shell mass and the counterterms
are replaced by their complex counterparts.
Moreover, renormalized complex masses are implicitly used in the calculation
of the self-energy.
In order to fix the counterterms, we generalize the renormalization
conditions of the complete on-shell scheme
\cite{Denner:1993kt,Aoki:1980ix} and require
\beqar \label{eq:ren-cond-CMS}
\hat\Si^{W}_{\rT}(\cmws) = 0, \qquad 
\hat\Si^{\prime W}_{\rT}(\cmws) = 0, \qquad
\eeqar
where the prime denotes differentiation with respect to the argument.
The first condition \refeq{eq:ren-cond-CMS} fixes the mass counterterm
in such a way that the renormalized W-boson mass is equal to the
location of the propagator pole in the complex plane. This is a
gauge-invariant quantity, as pointed out and shown in
\citeres{Stuart:1991xk,Sirlin:1991fd}. The second 
condition \refeq{eq:ren-cond-CMS} fixes the field
renormalization constant $\delta \cZ_{W}$.  Note that  $\delta \cZ_{W}$
exactly drops out in all $S$-matrix elements
that do not involve external W bosons, but allows to render all vertex
functions finite.  This generally holds for all field renormalization
constants of unstable particles in
$S$-matrix elements without external unstable particles.  Unlike in
\citeres{Denner:1993kt,Aoki:1980ix}, we did not take real parts in the
renormalization conditions \refeq{eq:ren-cond-CMS}, and thus not only
the mass renormalization constants but also the field renormalization
constants become complex. This ansatz is supported by the
fact that the imaginary part of one-loop scattering amplitudes
involving unstable external particles becomes gauge dependent if the
imaginary parts of the counterterms are not included
\cite{Denner:1997kq}.  

The renormalization conditions \refeq{eq:ren-cond-CMS} have the
solutions
\beqar\label{exact-complex-ren-const}
\de\cmws = \Si^{W}_{\rT}(\cmws), \qquad 
\de \cZ_{W} = - \Si^{\prime W}_{\rT}(\cmws), \qquad
\eeqar
which require to calculate the self-energies for complex squared
momenta $p^2=\cmws$. This would demand an analytic continuation of the
2-point functions entering the self-energies in the momentum variable
to the unphysical Riemann sheet. In order to avoid this complication,
we perform appropriate expansions about real arguments,
\beqar
\Si^{W}_{\rT}(\cmws) &=& \Si^{W}_{\rT}(\MW^2) +  (\cmws-\MW^2)
\Si^{\prime W}_{\rT}(\MW^2) \nl &&{}+ \Oaaa.
\eeqar
The $\Oaaa$ contributions result from products of terms
$\Si^{W}=\Oa$
and $\cmws-\MW^2=\Oa$
and are UV-finite by construction at the one-loop level.

By neglecting higher-order   
terms, that are beyond the accuracy needed for an $\Oa$ calculation,
we can replace \refeq{exact-complex-ren-const} by
\beqar\label{complex-ren-const}
\de\cmws &=& \Si^{W}_{\rT}(\MW^2)+  (\cmws-\MW^2)
\Si^{\prime W}_{\rT}(\MW^2) , \nl 
\de \cZ_{W} &=& - \Si^{\prime W}_{\rT}(\MW^2),
\eeqar
which corresponds to a slightly modified renormalization scheme.

When inserting 
\refeq{complex-ren-const} into
\refeq{eq:ren-se}, we can
rewrite the renormalized W-boson self-energy in the CMS as
\beq\label{eq:ren-se-cms}
\hat\Si^{W}_{\rT}(k^2) = \Si^{W}_{\rT}(k^2) - \de\MW^2
+(k^2-\MW^2)\de Z_{W}, 
\eeq
with 
\beq\label{complex-ren-const-ons}
\de\MW^2 = \Si^{W}_{\rT}(\MW^2), \qquad
\de Z_{W} = - \Si^{\prime W}_{\rT}(\MW^2),
\eeq
Equations \refeq{eq:ren-se-cms} with \refeq{complex-ren-const-ons}
have exactly the form of the renormalized W-boson self-energy in the usual
on-shell scheme, but without taking the real part of the counterterms.
While in the on-shell scheme the self-energies are calculated
with the real renormalized masses, in
\refeq{eq:ren-se-cms} and \refeq{complex-ren-const-ons} the
self-energies are to be calculated with the complex masses,
although with real squared momenta.  

The renormalization of the self-energies of the neutral gauge bosons,
of the Higgs boson, and of the massive and massless fermions is done in
the same spirit and can be found in \citere{Denner:2005fg}.

Owing to its definition \refeq {eq:compl-mixing}, the renormalization
of the complex weak mixing angle is given by
\beq\label{eq:ren-mixing-angle}
\frac{\de\csw}{\csw} = -\frac{\ccw^2}{\csw^2}\frac{\de\ccw}{\ccw}
=-\frac{\ccw^2}{2\csw^2}
\left(\frac{\de\cmws}{\cmws}-\frac{\de\cmzs}{\cmzs}\right).
\eeq

The electric charge is fixed in the on-shell scheme by requiring that
there are no higher-order corrections to the $ee\ga$ vertex in the
Thomson limit.  In the CMS this condition reads
\beq\label{eq:ren-charge}
\frac{\de e}{e} = \frac{1}{2}\Si^{\prime AA}(0) -
  \frac{\sw}{\cw}\frac{\Si^{AZ}_{\rT}(0)}{\cmzs}.
\eeq
Because of 
the complex masses in the loop integrals,
the charge renormalization constant $\de e$ and thus the renormalized
charge become complex. Since the bare charge is real, the imaginary
part of $\de e$ is directly fixed by the imaginary part of
self-energies. In a one-loop calculation, the imaginary part of the
renormalized charge drops out in the corrections to the absolute
square of the matrix element, because the charge factorizes from the
lowest-order matrix element.  Starting from the two-loop level, the
imaginary part contributes.

For a correct description of the resonances at the $\Oa$ level, we
need the width including $\Oa$ corrections. In the CMS the width is
implicitly defined via \refeq{complex-ren-const}. Using
$\de\cmws=M_{\PW,0}^2-\cmws$ and taking the imaginary part of the
first equation \refeq{complex-ren-const} yields
\beqar\label{width}
\MW\GW &=&\Im\{\Si^{W}_{\rT}(\MW^2)\}
\\&&{} 
- \MW\GW \Re\{\Si^{\prime W}_{\rT}(\MW^2) \} + \Oaaa,\nn
\eeqar
which can be iteratively solved for $\GW$. In $\Oaa$, \ie including
first-order corrections to the width, the result is equivalent to the
one obtained in the usual on-shell scheme. To this order the imaginary
part of the self-energy is required in two-loop accuracy, but the
$\Oa$-corrected width can be more easily obtained by calculating the
one-loop corrections to the decay processes $\PW\to \bar f f'$ 
in the usual on-shell renormalization scheme with a real W-boson mass.

The calculation of the width from \refeq{width} including $\Oaa$ terms
violates the relation between the width and the imaginary part of the
counterterm determined by \refeq{complex-ren-const}. 
It is not obvious that this procedure respects gauge invariance.
However, a change in the width can be compensated by a suitable change
in the mass counterterm, and an overall change of a one-loop
counterterm of a physical parameter does not violate gauge invariance
of a physical matrix element since it is obtained by a variation of
the gauge-invariant lowest-order matrix element.

In fact, gauge invariance of our result for $\Pep\Pem\to4\Pf$ has been
explicitly checked \cite{Denner:2005fg,Denner:2005es}
by performing the same calculation within the
background-field method \cite{Denner:1994xt}. In this framework, the
gauge-boson field renormalization constants can be determined in terms
of the parameter renormalization in such a way that Ward identities
keep their form upon renormalization.
The complex parameter renormalization is fixed
as in the conventional 't Hooft--Feynman gauge, and the concept of
complex renormalization can be applied in the same way as above.

\section{Summary}

The complex-mass scheme is a consistent, gauge-invariant scheme for the
calculation of higher-order corrections to processes with intermediate
unstable particles. It has been fully elaborated for one-loop
calculations and is straight-forward to apply. It has been used in the
calculation of the complete electroweak corrections to 
$\Pep\Pem\to4\,$fermions \cite{Denner:2005fg,Denner:2005es}
(via charged current) and to the Higgs-boson decay
$\PH\to4\,$leptons \cite{Bredenstein:2006rh}.

\end{document}